\documentclass[12pt,epsfig]{article}
\newcommand{\ba}{\begin{array}{c}}
\newcommand{\baz}{\begin{array}{cc}}
\newcommand{\bad}{\begin{array}{ccc}}
\newcommand{\bav}{\begin{array}{cccc}}
\newcommand{\ea}{\end{array}}

\usepackage{amssymb} 
\usepackage{epsfig}
\usepackage{float}
\newcommand{\be}{\begin{equation}}
\newcommand{\ee}{\end{equation}}
\newcommand{\bea}{\begin{eqnarray}}
\newcommand{\eea}{\end{eqnarray}}
\begin{document}

\begin{center}
\bf {Ambiguities pertaining to quark-lepton complementarity}
\end{center}

\begin{center}
C. Jarlskog 
\end{center}

\begin{center}
{\em Division of Mathematical Physics\\
LTH, Lund University\\
Box 118, S-22100 Lund, Sweden}
\end{center}

\begin{abstract}

Recently the possible origin of the so-called quark-lepton 
complementarity relations
has received a considerable amount of attention. We point out
some of the inherent ambiguities in such analyses.

\end{abstract}

\section{Introduction }

Recently the so-called Quark-Lepton Complementarity
(hereafter abbreviated by QLC)
has attracted much attention. The current status of this topic 
is eloquently described in a recent review by  
Minakata \cite{minakata} who gives a 
large number of references to relevant papers. 
Minakata, except in a few
cases, does not convey what each listed paper 
has actually contributed, an impossible task indeed. 
The main message is, however, crystal clear: 
the QLC is considered to be remarkable
and a large number of papers has been devoted to it.

First of all a brief account of what QLC is and why it
has attracted so much attention is in order. For a more detailed 
discussion the reader is again referred to Ref. \cite{minakata}.

The QLC relates
the angles of the quark mixing matrix to that of the lepton
mixing matrix. The empirical relation 
\begin{equation}
\theta_{12}^{quark} + \theta_{12}^{lepton} = \pi/4
\label{rel12}
\end{equation}
valid to within a couple of degrees is considered to be the
most remarkable of the two QLC relations. 
The second QLC relation reads 
\begin{equation}
\theta_{23}^{quark} + \theta_{23}^{lepton} = \pi/4
\label{rel23}
\end{equation}
This is not as interesting because $\theta_{23}^{quark}
$ is only about two degrees and therefore this equation would have been
satisfied, within the errors, even if this angle had been say
zero. Finally, a third
analogous relation
\begin{equation}
\theta_{13}^{quark} + \theta_{13}^{lepton} = \pi/4
\label{rel13}
\end{equation}
is badly violated as the sum of the angles
in the LHS is less than 10 degrees.

The reason for the popularity of QLC is that already since
three decades theorists have  
been almost desperately looking for some clues 
that may help them understand the pattern of masses
and mixings of fundamental fermions. All models
proposed so far for this purpose have severe drawbacks,
either by containing too many assumptions or free parameters. 
A hint from
experiments would be very welcome indeed. The empirical values
of the quark masses and the elements of the quark mixing
matrix have been known for quite sometime. Yet we have
failed to produce a reasonable model which could explain
them. The corresponding lepton sector has exhibited a
remarkably different pattern, with large mixing angles.
It would be no less than fantastic if neutrino oscillation
measurements could provide the solution to the long-standing
problem of masses and mixings in
the quark sector. It is argued that QLC may 
actually serve such a purpose.

Evidently any measurement
could have a message or more to convey. But how these
messages are to be interpreted is not always trivial. 
In this article we emphasise a few of the
uncertainties that could easily invalidate the QLC analyses.
The point of this paper is by no means to discourage 
the reader from working on
the eminently fundamental problem of masses and mixings
of fundamental fermions but to bring to
his/her attention some of the pitfalls so that this
outstanding topic is not 
degraded to the level of "playing QLC games".    

\section{Ambiguities in QLC}

A straight-forward issue of concern has to do with 
mixing angles and phases. These are by no
means "invariant" quantities and therefore 
their definitions, signs and
magnitudes are convention dependent. 
Not even the conventions adopted by the PDG \cite{pdg}
may be taken as sacred.
As an example,
the CP phase in one convention generally contains mixing
angles of a different convention. The invariants of
the quark mixing matrix are known to be the 
magnitudes of the elements of that matrix. Even the
quantity $J$, introduced as a measure of CP violation, is a function
of these magnitudes as it is twice the area of any of the
six unitarity triangles \cite{stora88}. Suppose for a moment
that the  
parameterisation of the quark and lepton mixing matrices
had been unique. Even then 
there would still be ambiguities when we add angles from
these two different sectors. As an example one may read the
message from experiments on 
$\theta_{12}^{quark} + \theta_{12}^{lepton}$ as
being $12+33 =45$ degrees but equally well as 
$-12+33 =21$ degrees. Additional phases, such as
Majorana phases of neutrinos, would give us other
possible interpretations.

Another serious source of theoretical error 
has to do with the fact that with
massive neutrinos lepton-hadron universality is 
generally lost in a substantial way.
In other words, the leptonic sector is 
not a copy of the quark sector.

The simplest way to accommodate massive neutrinos
is to keep the symmetry as well as the family structure 
of the Minimal Standard Model and to introduce a number of 
passive (right-handed) neutrinos
($\nu_R$'s). The neutrino mass matrix is then of the 
familiar form
\begin{equation}
\cal{M_\nu} = \left(
\begin{array}{cc}
0 & m_D \\
{m}_D^T & M_R
\end{array}
\right)
\end{equation}
where $m_D$ and $M_R$ are the so-called Dirac and Majorana
mass matrices and the $T$ stands for transposition. With
three left-handed and $n$ right-handed neutrinos
the matrix $m_D$ is 3-by-n and $M_R$ is n-by-n.
Furthermore, the matrix $M_R$ 
is symmetric by construction but otherwise arbitrary. 
There is no justification for taking 
$M_R$ to be diagonal. However,
one can start by diagonalising $M_R$ through 
\begin{equation}
M_R = U_R D_R {U}^T_R
\label{U_R}
\end{equation}
where $U_R$ is the required unitary matrix such that $D_R$ is
diagonal. With three right-handed neutrinos one has
\begin{equation}
D_R = \left(
\begin{array}{ccc}
M_1 &0 & 0\\
0& M_2 & 0\\
0&0 & M_3
\end{array}
\right)
\label{diagmajor}
\end{equation}
where the $M_j$, j=1-3 are real parameters.
The mass matrix can now be written in the form
\begin{equation}
\cal{M_\nu} = \left(
\begin{array}{cc}
1 & 0 \\
0 & U_R
\end{array}
\right)
\left(
\begin{array}{cc}
0 & m_D U^\star_R \\
{U}^\dagger_R {m}^T_D & D_R
\end{array}
\right)
\left(
\begin{array}{cc}
1 & 0 \\
0 & {U}^T_R
\end{array}
\right)
\end{equation}
and the matrices $U_R$ and ${U}^T_R$ can be absorbed into
the definition of the right-handed neutrinos. For a detailed
discussion of this point see, for example, Ref. \cite{ceja91}. 
However $U_R$ 
leaves its mark by modifying the definition of the 
Dirac mass matrix
\begin{equation}
m_D \rightarrow m^\prime_D = m_D U^\star_R
\end{equation}
$m^\prime_D$ being the effective Dirac mass matrix
and as such
it contains parameters stemming from the right-handed sector. 
Since the matrix $U_R$ is unknown to us there is
no justification in assuming that the resulting effective Dirac mass
matrix would have any resemblance to the mass matrices of the
quarks. 

In the see-saw type models, the $M_i$, in 
Eq.(\ref{diagmajor}), are assumed to be huge as compared to the
weak scale. In such scenarios
the effective three-by-three
neutrino mass matrix for the light neutrinos is given by
\begin{equation}
m_\nu = -{m}^{\prime T}_D ~{1 \over D_R}~ m^\prime_D
\end{equation}

To further elucidate this point we now consider 
a very simple case where $M_3$ is much
larger than the other two mass scales $M_1$ and $M_2$ so
that we can neglect $1/M_3$. We denote the three columns of the
effective Dirac mass matrix, which we take to be real, by $\vec{a}$, 
$\vec{b}$ and $\vec{c}$ respectively, 
\begin{equation}
m^\prime_D \equiv
\left( \begin{array}{ccc} \vec{a}& 
\vec{b} & \vec{c} \end{array} \right)
\end{equation}
Using this notation the neutrino mass matrix, up to an overall
normalisation factor, is given by
\begin{equation}
m_\nu = \left( \begin{array}{c} \vec{a}^T \\
\vec{b}^T \\ \vec{c}^T \end{array} \right)
\left( \begin{array}{ccc} 1& 0 & 0 \\
0 & r & 0 \\  0 & 0 & 0 \end{array} \right)
\left( \begin{array}{ccc} \vec{a}& 
\vec{b} & \vec{c} \end{array} \right)
\end{equation}
Here $r = M_1/M_2$, is a free parameter.
The eigenvalues, $\lambda$, of this matrix are  
solutions of the equation
\begin{equation}
\lambda^3 - \lambda^2 tr m_\nu + {\lambda \over 2} \{(tr m_\nu)^2
-tr (m_\nu^2)\} - det m_\nu = 0
\label{chareq}
\end{equation} 
Since the determinant of $m_\nu$ vanishes one of the eigenvalues is
zero. The third column of the effective Dirac mass matrix
thus decouples and we can write the mass matrix in the following
very simple form
\begin{equation}
m_\nu = \vert a > < a \vert + r \vert b> < b \vert
\label{abmatrix}
\end{equation}
where $(\vert a > < a \vert)_{jk} \equiv a_j a_k$, etc. Note also
that one could absorb the parameter $r$ into the definition of
the vector $b$, if one so wishes. 
The traces in Eq.(\ref{chareq}) are easily deduced  
from Eq.(\ref{abmatrix}) and
the corresponding formula for the square of the mass matrix, i.e.,
\begin{equation}
m_\nu^2 = a^2 \vert a > < a \vert + r \vec{a} . \vec{b}~  
\{\vert a >< b \vert + \vert b >< a \vert \} + r^2 b^2 \vert b >< b \vert
\end{equation}
where $a^2 = \vec{a} . \vec{a}$ and $b^2 = \vec{b} . \vec{b}$.
We find 
\begin{equation}
tr m_\nu = a^2 + r b^2
\end{equation}
and
\begin{equation}
tr m_\nu^2 = a^4 + 2r (\vec{a} . \vec{b})^2 + r^2 b^4
\end{equation}
Thus the non-zero eigenvalues are given by
\begin{equation}
\lambda_{\pm} = {1 \over 2}\{a^2 + r b^2 \pm 
\sqrt{ ({a^2 + r b^2})^2 -4 r \vert \vec{a} \times 
\vec{b} \vert^2} ~ \}
\end{equation}
To find the matrix $V$ that diagonalises $m_\nu$ we must compute
the corresponding eigenvectors.

As an example, let us consider the case where 
vectors $a$ and $b$ are orthogonal to
each other. In this case, the two non-zero eigenvalues
are given by
\begin{equation}
\lambda_+ = a^2, ~~~\lambda_- = r b^2
\end{equation} 
and the eigenvectors are unit vectors along
$\vec{a}$, $\vec{b}$ and $\vec{a} \times \vec{b}$ respectively.
Thus we have
\begin{equation}
V =
\left( \begin{array}{ccc}
{a_1 \over a} & {b_1\over b}& {a_2 b_3 - a_3 b_2 \over ab} \\
{a_2 \over a} & {b_2\over b}& {a_3 b_1 - a_1 b_3 \over ab} \\
{a_3 \over a} & {b_3\over b}& {a_1 b_2 - a_2 b_1 \over ab}
\end{array}
\right)
\end{equation}
If we wish to reproduce the popular bimaximal mixing matrix
\cite{minakata} and
none of the components of vectors $a$ and $b$ are zero
we can take one of the components of the cross product
to be zero. Taking the first component to be zero gives
\begin{equation}
{a_3 \over a_2} = {b_3 \over b_2} \equiv \alpha
\end{equation}
The matrix $V$ is then given by 
\begin{equation}
V =
\left( \begin{array}{ccc}
cos\phi_1 & - sin\phi_1& 0 \\
sin\phi_1 cos\phi_2 & cos\phi_1 cos\phi_2 & -sin\phi_2 \\
sin\phi_1 sin\phi_2 & cos\phi_1 sin\phi_2& cos\phi_2 
\end{array}
\right)
\label{phi12}
\end{equation}
where
\begin{equation}
cos\phi_1 = {a_1 \over a}, ~sin\phi_1 = -{b_1 \over b}
\end{equation}
\begin{equation}
cos\phi_2 = {1 \over \sqrt{1+\alpha^2}}, ~sin\phi_2 = 
{-\alpha \over \sqrt{1+\alpha^2}}
\end{equation}
The matrix in Eq.(\ref{phi12}) can also be written in the form
\begin{equation}
V =
\left( \begin{array}{ccc}
1 & 0 & 0 \\
0 & cos\phi_2 & -sin\phi_2 \\
0 & sin\phi_2& cos\phi_2 
\end{array}
\right)~
\left( \begin{array}{ccc}
cos\phi_1 & - sin\phi_1& 0 \\
sin\phi_1 & cos\phi_1  & 0 \\
0 & 0& 1
\end{array}
\right)
\end{equation}
For $\phi_1 = \phi_2 = \pi /4$ this matrix is exactly 
the bimaximal matrix discussed by many authors (see \cite{minakata}).
By putting $\phi_1 = \pi /6$ and $\phi_2 = \pi/4$ we
obtain the currently popular matrix
\begin{equation}
V =
\left( \begin{array}{ccc}
{\sqrt{3} \over 2} & -{1 \over 2}& 0 \\
{1 \over 2 \sqrt{2}} & {\sqrt{3} \over 2 \sqrt{2}}  & - {1 \over  \sqrt{2}}\\
{1 \over 2 \sqrt{2}} & {\sqrt{3} \over 2 \sqrt{2}}& {1 \over  \sqrt{2}}
\end{array}
\right)
\end{equation} 
as discussed, for example, by Petcov \cite{petcov}.
Furthermore, by tuning the parameter $r$ 
we may change the mass-splitting of the two massive
neutrinos and by adjusting $a$ we may
vary the scale of these masses. Thus by making assumptions
we can get the required results. There is no sign that the Dirac
mass matrices of the quarks play any role here.

\section{Conclusions}  

In this article we have pointed out that in general
adding mixing angles from the quark and lepton sectors
does not make sense. Furthermore, 
not knowing the right-handed Majorana mass matrix $M_R$ 
makes it impossible 
to relate the quark and lepton mixing matrices. Any assumption
about $M_R$ is a source of uncertainty which easily
invalidates the subsequent conclusions.

A further uncertainty arises as the matrix $V$, discussed above 
is still not the
lepton mixing matrix. It has to be multiplied with
the (to us unknown) unitary matrix that diagonalises 
the mass matrix of charged leptons.

Note that the hypothesis that leptogenesis is responsible for the
observed baryon asymmetry of the universe brings $M_R$, at least
partially, into the realm of observability. However, current models
contain many assumptions and free parameters (see, for example,
a recent review by T. Hambye \cite{hambye}). The upshot of
these studies is that leptogenesis looks very promising, as
a mechanism for generating the observed baryon asymmetry of the
universe. But the specific models are far from being 
sufficiently pinned-down to be useful for connecting quark
and lepton mixing angles as is done in QLC. Further progress
in this domain would be highly welcome.

\end{document}